\documentclass{aa}
\usepackage[dvips,xdvi]{graphics}

\def\BA{\mbox{Br$\alpha$}}
\def\HB{\mbox{H$\beta$}}
\def\AV{\mbox{A$_{\rm V}$}}

\def\L{\mbox{$\lambda$}}

\def\Ne{\mbox{$n_e$}}

\def\Te{\mbox{$T_e$}}

\begin{document}

\title{
Infrared spectroscopy of young supernova remnants
heavily interacting with the interstellar medium 
% Far infrared spectroscopy of the young supernova remnant RCW103
\thanks{ Based on observations with ISO, an ESA project
with instruments funded by ESA Member States (especially the PI countries:
France, Germany, the Netherlands and the United Kingdom)
and with the participation of ISAS and NASA. 
%The SWS is a joint project of SRON and MPE.
}}
\subtitle{ 
I. Ionized species in RCW103 }

\author{E. Oliva\inst{1}, 
A.F.M Moorwood\inst{2},
S. Drapatz\inst{3},
D. Lutz\inst{3}, and
E. Sturm\inst{3}
}

\institute{
Osservatorio Astrofisico di Arcetri, Largo E. Fermi 5, 
I--50125 Firenze, Italy
\and
European Southern Observatory, Karl Schwarzschild Str. 2, D-85748
Garching bei M\"unchen, Federal Republic of Germany 
\and
Max Planck Institute f\"ur Extraterrestrische Physik,
            Postfach 1603, D-85740 Garching, Germany
}

\offprints{E. Oliva}

\date{Received 16 October, accepted ... December 1998 }

\titlerunning{ISO infrared spectroscopy of fast shocks. I. 
Ionized lines in RCW103}
\authorrunning{Oliva et al. }
\thesaurus{08(                  % Diffused matter in space
               02.01.2;          % Atomic data
               09.01.1;          % ISM: abundances
               09.09.1 RCW103;   % ISM: individual objects
               09.19.2;          % ISM: supernova remnants
               11.19.2;          % Galaxies: Seyfert
	       13.09.4           % Infrared: ISM: lines and bands
             )}
\maketitle

\begin{abstract}

ISO spectral observations
of the supernova remnant RCW103 are presented. This object
is the prototype of relatively
young remnants ($\sim\!10^3$ yr) with
fast shocks ($v_s\!\sim\!1000$ km/s) 
interacting with dense interstellar medium.
%Its spectrum could therefore be useful 
%as a template for the study of complex objects, 
%e.g. active galaxy nuclei,
%where the contribution of shock excited gas is unclear.
%

The spectrum is dominated by prominent lines of [NeII], [SiII],
[FeII] and other low excitation species
which provide, for the first time, a  simple and reliable estimate
of the gas abundances of refractory (Si, Fe, Ni) and non--refractory
(Ne, P, S, Cl, Ar) species. Apart from nickel, all the derived abundances are 
close to solar,
confirming that the shock has destroyed all dust grains. 
Like the optical nickel lines, [NiII]\L6.64 $\mu$m yields Ni
abundances a factor $\simeq$10 solar which we propose results from
a large underestimation of the computed Ni$^+$ collision strengths.

The observed intensities and velocity widths of ionic lines are compatible
with emission from the post--shock region alone with only a very small
(if any) contribution from the photoionized precursor.
This result does not agree with shock models which
predict that the precursor should emit powerful line emission, especially
from highly ionized species.
The possible consequence of this on the modelling of Seyfert spectra
is briefly discussed.

\end{abstract}
\keywords{
Atomic data; 
ISM: abundances; ISM: supernova remnants; 
Galaxies: Seyfert;
Infrared: ISM: lines and bands
}

\section{Introduction}

Radiative supernova remnants 
are the ideal astrophysical laboratory for studying the
emission spectrum of relatively fast shocks interacting with
the interstellar medium
(e.g. Draine \& McKee \cite{drainemckee}).
In the standard, idealized, view the supernova blast wave expands through a
uniform, low density ($\approx$1 cm$^{-3}$) medium and the shock remains
adiabatic, i.e. the postshock gas has insufficient time to cool/recombine,
for $>\!10^4$ yr and up until the shock has slowed
to $v_s\!<\!200$ km/s. After this the shock becomes radiative with the
postshock region recombining and radiating away most of
the shock mechanical energy in the form of UV, optical and IR lines
whose surface brightness simply scale with the shock energy flux
$n\,v_s^3$, $n$ being the pre--shock density.
In practice, a `standard' radiative supernova remnant is expected to emit
weak lines, i.e. $\Sigma(\HB)\!\la\!10^{-5}$ erg cm$^{-2}$ s$^{-1}$ sr$^{-1}$
(cf. Sect 3.2 of Dopita \& Sutherland 1996, hereafter \cite{DS96}) and 
2--3 orders
of magnitude weaker than those observed in bright supernovae remnants
such as RCW103 (Oliva et al. 1989, hereafter \cite{OMD89}). This simple fact
indicates, therefore, that the shock of RCW103 is much faster
or, equivalently, that the SNR is much younger than the canonical values.
This is also confirmed by X--ray spectra which yield an age of only
$10^3$ yr and a main shock velocity of about 1200 km/s
(Nugent et al. \cite{nugent}). 
\begin{figure*}
%\centerline{\resizebox{\hsize}{!}{\rotatebox{-90}{\includegraphics{sws02.ps}}}}
\centerline{\resizebox{17.3cm}{!}{\rotatebox{-90}{\includegraphics{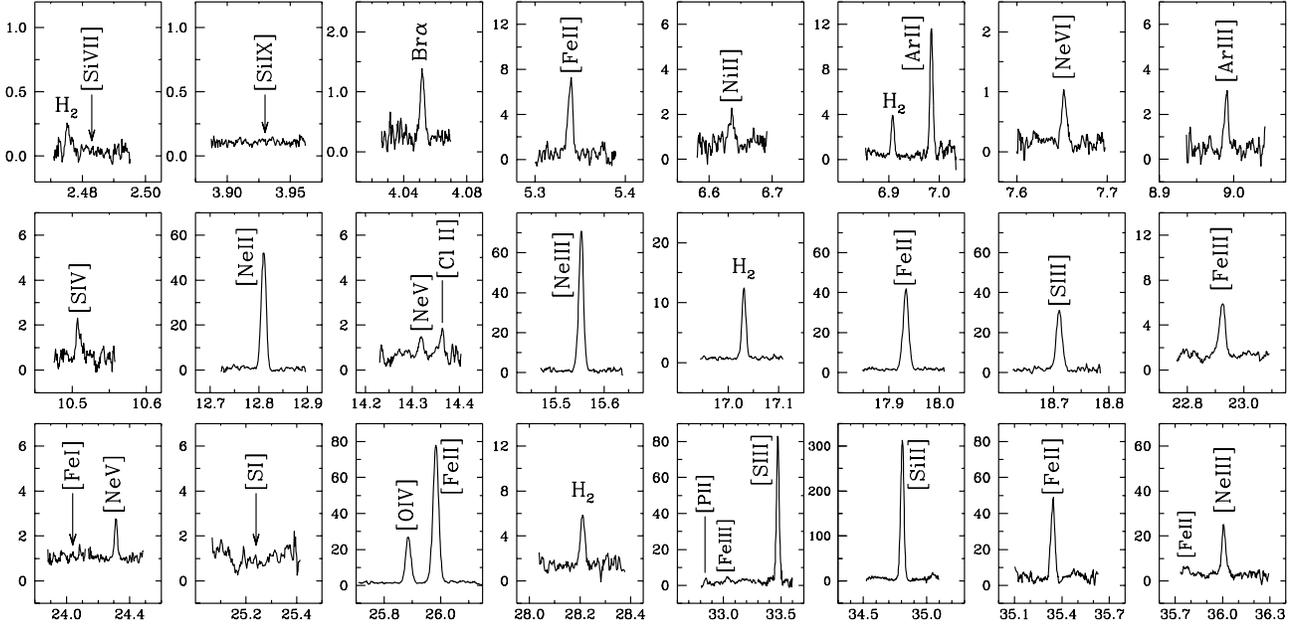}}}}
\vskip-5pt
\caption{
Individual line scans with SWS02. Wavelengths are in $\mu$m and fluxes
in Jy. Arrows mark the positions of undetected lines.
}
\label{sws02}
\end{figure*}

\begin{figure}
\centerline{\resizebox{\hsize}{!}{\rotatebox{0}{\includegraphics{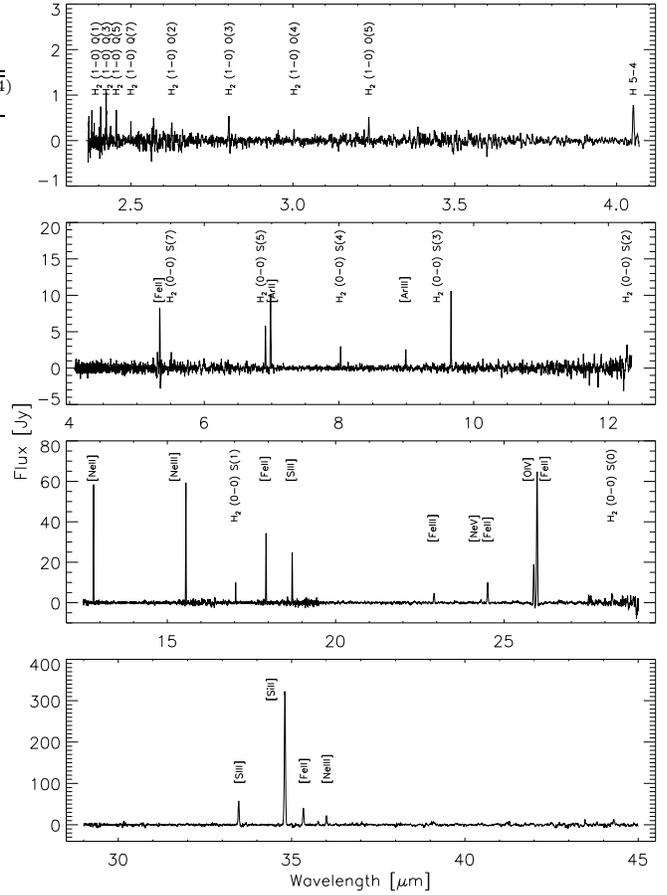}}}}
\vskip-5pt
\caption{
Complete 2.4-45 $\mu$m SWS01 spectrum with line identification.
Note that each spectral segment has been ``continuum subtracted''
to remove instrumental offsets and drifts.
The continuum level can be estimated from the PHT-S spectrum
(Fig.~\ref{phts}) and from the deeper SWS02 line scans (Fig.~\ref{sws02}).
}
\label{sws01}
\end{figure}

\begin{figure}
\centerline{\resizebox{\hsize}{!}{\rotatebox{90}{\includegraphics{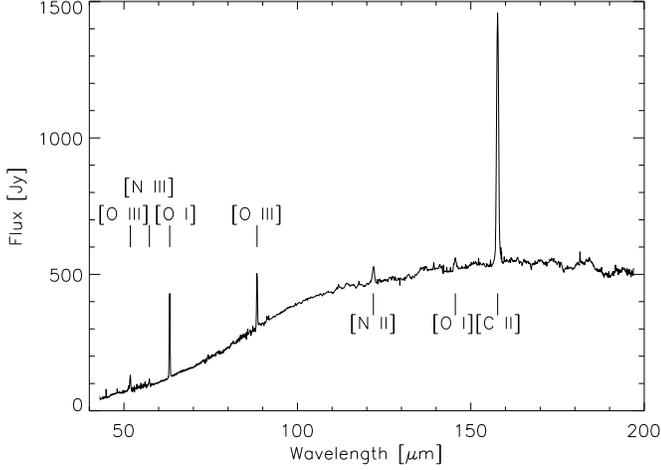}}}}
\vskip-5pt
\caption{
Complete LWS01 spectrum
 with lines identification. Note that 
RCW103 lies close to the the Galactic plane ($b=-0.36$). Therefore,
the 100~$\mu$m continuum and
the [OI], [CII], [NII] lines are strongly contaminated, and most probably
dominated by fore/background emission. 
%See text, Sect.~\ref{observations} for details.
}
\label{lws}
\end{figure}

\begin{figure}
\centerline{\resizebox{\hsize}{!}{\rotatebox{0}{\includegraphics{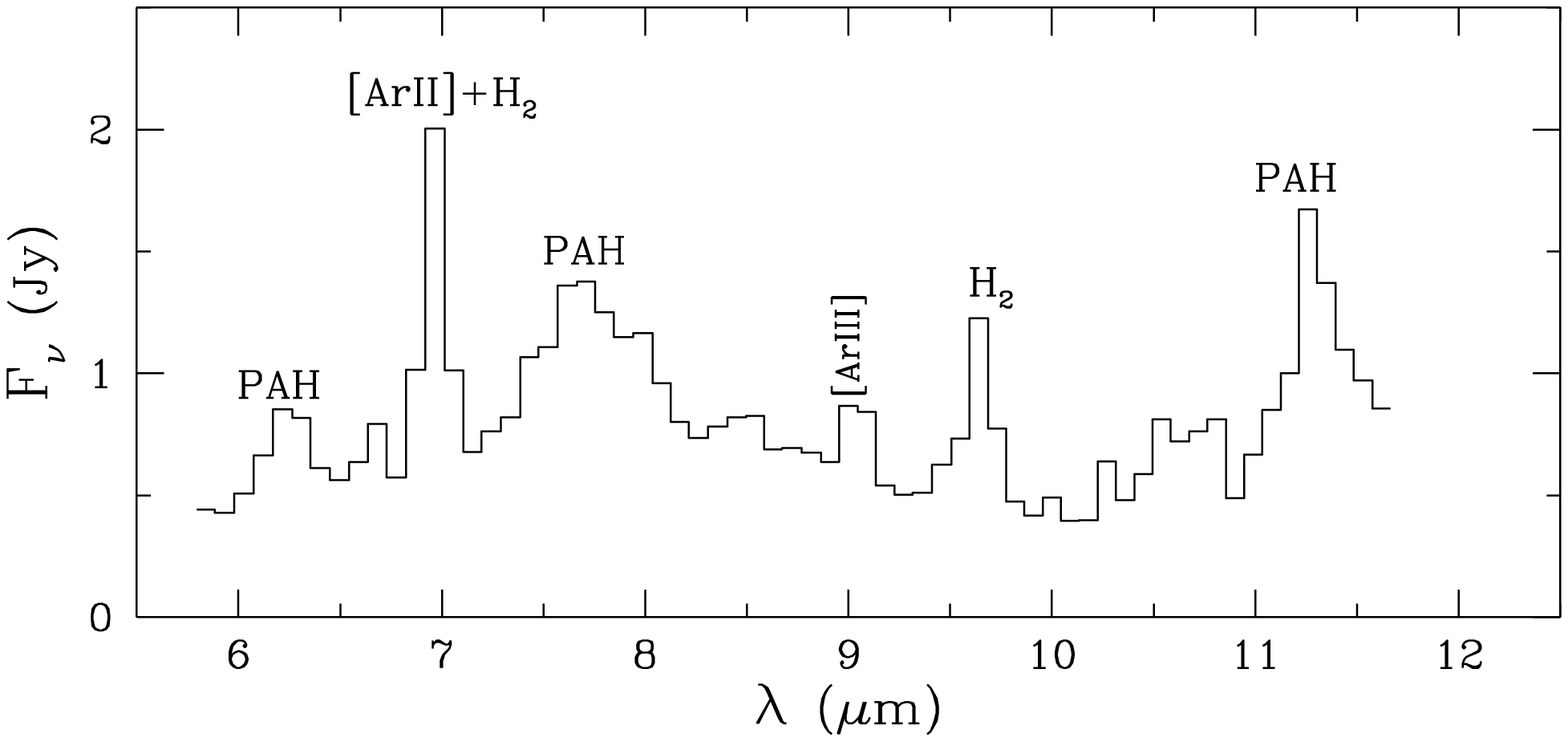}}}}
%\centerline{\resizebox{7.0cm}{!}{\rotatebox{0}{\includegraphics{phts.ps}}}}
\vskip-5pt
\caption{
Complete PHT-SL spectrum. 
The PAH features are most probably dominated
by fore/background emission from the Galaxy disk.
}
\label{phts}
\end{figure}
Thus, the prominent line emission from the optical filaments of RCW103
results from the impact of the $v_s\!\sim\!1000$ km/s SNR blast wave 
onto dense clouds of ISM material (probably molecular clouds). 
Due to the
higher density, the postshock gas  cools and recombines much faster than
in the canonical model and the SNR becomes radiative at much
earlier times. The large surface brightness of the emitted line
spectrum simply reflects the large mechanical energy flux of the shock.

Studying the IR spectrum of radiative shocks can give interesting 
information on the metal abundances and, by comparing
refractory and non--refractory species,
can yield a direct measurement of the actual efficiency of grain
destruction by the shock.
Moreover, the spectra can be used to verify the predictions of models
of fast shocks, which have been recently published by \cite{DS96}.

Being the brightest and best studied SNR, RCW103 was chosen as the 
prototype object for ISO template spectra of gas excited by fast shocks.
The observations are described in Sect.~\ref{observations} and the
derived physical parameters are discussed in Sect.~\ref{physical_param}.
The results are compared with shock model predictions in 
Sect.~\ref{discussion} where we
draw particular attention to the relative roles
of the photoionized precursor and post--shock region in producing the
observed lines.
%In Sect.~\ref{conclusions} we draw our conclusions.

\section { Observations and results }
\label{observations}

\subsection{ ISO spectroscopy }
\label{iso_obs}

\begin{table}
\caption{Observed ISO line fluxes }
\label{tab_flux}
\def\SKIP#1{\noalign{\vskip#1pt}}
\def\P{$\,$(}
\def\UNC{\rlap{:}}
\def\UNO{\rlap{$^{(1)}$}}
\def\DUE{\rlap{$^{(2)}$}}
\def\TRE{\rlap{$^{(3)}$}}
\def\QUA{\rlap{$^{(4)}$}}
\def\NB{\rlap{$^a$}}
\def\NC{\rlap{$^b$}}
\def\ND{\rlap{$^c$}}
\def\NE{\rlap{$^d$}}
\def\X{ $\times$ }
\def\SB{$\rm [$}
\begin{flushleft}
\begin{tabular}{lcccc}
\hline\hline
\SKIP{1}
 Line & \hglue00pt Flux\UNO &  A$_\lambda$\DUE &
            \hglue10pt   Slit\TRE\hglue10pt\  & Aper. corr.\QUA\\
\SKIP{2}
\hline
\SKIP{1}
%
% line   F-SWS02 and  F-SWS01 units of 1e-13 erg cm-2 s-1 = 1e-20 W cm-2

%H$_2$ (1,0)Q(1) 2.406 & 7.4\NB & -- &14\X20 & -- \\ % se vogliamo agganciarci al NIR
%H$_2$ (1,0)Q(5) 2.454 & 7.7\NB & -- & 14\X20 & -- \\ % se vogliamo agganciarci al NIR
%H$_2$ (1,0)Q(6) 2.626  &  1.7 & --  & 14\X20 & -- \\
\SB SiVII]\L2.483   & $<$1.2 & 0.38  & 14\X20 & 1.0 \\
\SB SiIX]\L3.936    & $<$0.5 & 0.17  & 14\X20 & 1.0 \\
Br$\alpha$ \L4.051      &  7\P1) & 0.16  & 14\X20 & 1.0 \\  
%\SB FeII]\L5.340    &  57\P17)\NA   & 0.10 & 14\X20 & 1.0 \\ % SWS01=75, SWS02=45
\SB FeII]\L5.340    &  57\P17)   & 0.10 & 14\X20 & 1.0 \\ % SWS01=75, SWS02=45
%H$_2$ (0,0)S(7) \L5.509  & -- & 14\NB\UNC  & 14\X20 & -- \\  % Only SWS01 value
\SB NiII]\L6.635    &   6\P1.5)  & 0.07 & 14\X20 & 1.0 \\
%H$_2$ (0,0)S(5) 6.908  & 16 & --   & 14\X20 & -- \\ % SWS01=31 My SWS01 gives 34
%\SB ArII]\L6.984    &  62\P12)\NA & 0.06  & 14\X20 & 1.0 \\ % SWS01=72, SWS02=51
\SB ArII]\L6.985    &  62\P12) & 0.06  & 14\X20 & 1.0 \\ % SWS01=72, SWS02=51
\SB NiIII]\L7.349   & $<$3 & 0.06  & 14\X20 & 1.0 \\
Pf$\alpha$ \L7.458       & $<$3 & 0.06 & 14\X20 & 1.0 \\
\SB NeVI]\L7.655    &  1.6\P.3)  & 0.06& 14\X20 & 1.0 \\
%H$_2$ (0,0)S(4) 8.026  &  11\NB & -- & 14\X20 & -- \\  % Only SWS01 value
%\SB ArIII]\L8.991   &  7\P1)\NA  & 0.04 & 14\X20 & 1.0 \\ % SWS01=8 SWS02=6
\SB ArIII]\L8.990   &  7\P1)  & 0.04 & 14\X20 & 1.0 \\ % SWS01=8 SWS02=6
%H$_2$ (0,0)S(3) 9.665  &  28\NB  & --& 14\X20 & --\\ % Only SWS01 value
\SB SIV]\L10.51     &  2\P.5)  & \llap{$\sim$}0.5\NC & 14\X20 & 1.0 \\
%\SB NiI]\L11.30    &   $<$3   & -- & 14\X20 & 1.0 \\ % Serve?
%H$_2$ (0,0)S(2) 12.28  &  8   & -- & 14\X20 & --  \\
\SB NeII]\L12.81    & 120\P10)  & -- & 14\X27 & .85 \\
%\SB MgV]\L13.5     &  $<$1.5 & --   & 14\X27 & .85 \\% serve?
\SB NeV]\L14.32     &  1.3\P.3)  & --  & 14\X27 & .85 \\
\SB ClII]\L14.36    &  1.7\P.4)  & -- & 14\X27 & .85 \\
\SB NeIII]\L15.56   &  107\P10)  & -- & 14\X27 & .85 \\
%H$_2$ (0,0)S(1) 17.03  &  11 & --  & 14\X27 & -- \\ % SWS01=12
\SB FeII]\L17.93    &  51\P6) & -- & 14\X27 & .85 \\ 
\SB SIII]\L18.71    &  35\P4)  & -- & 14\X27 & .85 \\ 
% \SB FeVI]\L19.5   & $<$7 & -- & 14\X27 & .85 \\ %Serve?
\SB FeIII]\L22.92   &  8\P1)  & -- & 14\X27 & .85 \\ 
\SB FeI]\L24.04     &  $<$1  & -- & 14\X27 & .85 \\
\SB NeV]\L24.32     &  1.9\P.3)   & -- & 14\X27 & .85 \\
\SB FeII]\L24.51    &  13\P3)\NB & -- & 14\X27 & .85 \\
\SB SI]\L25.24      &  $<$1   & -- & 14\X27 & .85 \\
\SB OIV]\L25.88     &  27\P3)   & -- & 14\X27 & .85 \\ %SWS01=25
\SB FeII]\L25.98    &  90\P10)  & -- & 14\X27 & .85 \\ 
%H$_2$ (0,0)S(0) 28.21  & 3.6 & --  & 20\X27 & -- \\
\SB PII]\L32.8      &  2\P.6) & --  & 20\X33 & .58 \\ % marginally detected
\SB FeIII]\L33.0    &  2\P.6) & --  & 20\X33 & .58 \\ % marginally detected
\SB SIII]\L33.47    & 57\P6)   & -- & 20\X33 & .58 \\  
\SB SiII]\L34.8      & 280\P40) & -- & 20\X33 & .58 \\ % SWS01=300, my SWS01=310
\SB FeII]\L35.34      & 36\P4) & --  & 20\X33 & .58 \\ 
\SB FeII]\L35.77      &  3\P.8) & --  & 20\X33 & .58 \\ 
\SB NeIII]\L36.01     & 15\P3)  & -- & 20\X33 & .58 \\ % SWS01=17 from SWS01 I get 16
\SKIP{5}
\SB OIII]\L51.7       &  190\P30) & -- & $\oslash$ 80 & -- \\
\SB NIII]\L57.3       &   57\P14) & -- & $\oslash$ 80 & -- \\
\SB OI]\L63.2\ND      &  710\P80)\ND  & -- & $\oslash$ 80 & -- \\
Unidentified \L74.2\NE & 22\P7)  & -- & $\oslash$ 80 & -- \\ % Si mette o no?
\SB OIII]\L88.2       &  220\P30) & -- & $\oslash$ 80 & -- \\
\SB NII]\L121\ND      &  75\P13)\ND  & -- & $\oslash$ 80 & -- \\
\SB OI]\L145\ND      &  33\P7)\ND  & -- & $\oslash$ 80 & -- \\
\SB CII]\L157\ND      &  630\P70)\ND  & -- & $\oslash$ 80 & -- \\
\SKIP{2}
\hline
\SKIP{2}
\end{tabular}
\def\NOTA#1#2{
\hbox{\vbox{\hbox{\hsize=0.030\hsize\vtop{\centerline{#1}}}}
      \vbox{\hbox{\hsize=0.97\hsize\vtop{\baselineskip2pt #2}}}}\vskip2pt}
\NOTA{ $^{(1)}$ }{ Observed line flux, units of $10^{-20}$ W cm$^{-2}$,
errors are given in parenthesis }
%a column denotes uncertain values }
\NOTA{ $^{(2)}$ }{ Extinction (mag) extrapolated from A$_{\rm H}$=0.75
(\cite{OMD90}) using A$_\lambda\!\approx\!\lambda^{-1.7}$ up to 9 $\mu$m.
Reddening is assumed to be negligible beyond 12 $\mu$m}
\NOTA{ $^{(3)}$ }{ Size (in arcsec) of the SWS ($\lambda\!<\!45$ $\mu$m) and
LWS ($\lambda\!>\!45$ $\mu$m) apertures }
\NOTA{ $^{(4)}$ }{ Correction factor to account for different apertures,
based on the IRSPEC map of [FeII]\L1.644 (see caption of Fig.~\ref{irspec}).
% and valid only for ionized species
}
%\NOTA{ $^a$ }{ SWS02 and SWS01 measurements differ by more than 20\% }
\NOTA{ $^a$ }{ Only SWS01 measurement available }
\NOTA{ $^b$ }{ Adopting a ``standard'' 
A(silicate feature)/A$_{\rm V}\!\simeq\!0.1$
ratio }
\NOTA{ $^c$ }{ Line flux is probably contamined by back/foreground galactic
emission}
\NOTA{ $^d$ }{ Unidentified feature also seen in NGC7027 (Liu et al.
\cite{liu}) }
\end{flushleft}
\end{table}

Complete SWS01 (speed 4, 6700 sec total integration time) and LWS01
(4400 sec total integration time, no background spectrum subtracted)
spectra centered at the peak of [FeII]1.644 $\mu$m line emission
(cf. Fig.~\ref{irspec}) were obtained on February 6 and February 20, 1996.
These were complemented by a quick PHT-S spectrum (448 sec, February 20)
and deeper SWS02 observations at selected wavelengths
(total 7000 sec, obtained on August 15, 1996 and February 17, 1997),
always centered at the same position. The short wavelength section
(2.5--5 $\mu$m) of the PHT-S spectrum was strongly contaminated by
detector memory effects (a very bright source was observed
just before RCW103), the PHT-SS results are therefore unreliable and not
presented here.

The SWS data were reduced using standard routines of the SWS interactive
analysis system (IA) using calibration tables as of September 1997.
Reduction relied mainly on the default pipeline steps, plus
removal of signal spikes, elimination of the most noisy
band 3 detectors, and flat--fielding.
The LWS spectrum is
based on the end--product of the automatic pipeline as of April 1997 (i.e.
OLP 6).
A post-processing was performed within the ISO Spectral Analysis Package
(ISAP)\footnote{The ISO Spectral Analysis Package (ISAP) is a joint
development by the LWS and SWS Instrument Teams and Data Centers.
Contributing institutes are CESR, IAS, IPAC, MPE, RAL and SRON.},
Version 1.2,
with special emphasis on removal of signal spikes and memory effects,
averaging of the different scans, and flat-fielding of the 10 detectors.
 
\begin{figure*}
  \resizebox{\hsize}{!}{\rotatebox{0}{\includegraphics{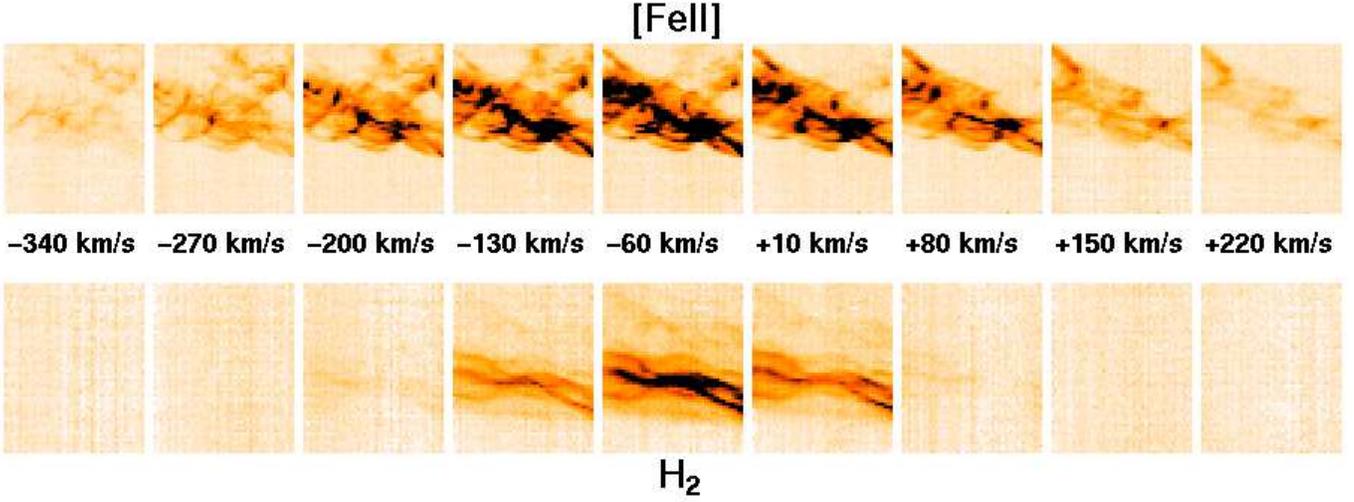}}}
\caption{
Line images at various velocity bins, reconstructed from the IRSPEC spectra
discussed in Sect.~\ref{observations}, 
%
% Modified
% 
all frames are cut to the same levels as the $v$=--60 km/s images.
Note that [FeII] (1.644 $\mu$m)
shows bright filaments moving by up to $\pm$250 km/s, 
while H$_2$ (2.121 $\mu$m)
is narrow and unresolved (i.e. FWHM$<$130 km/s).
The colour images are also available at
http://www.arcetri.astro.it/$\sim$oliva
}
\label{velcuts}
\end{figure*}

\begin{figure}
  \resizebox{\hsize}{!}{\rotatebox{0}{\includegraphics{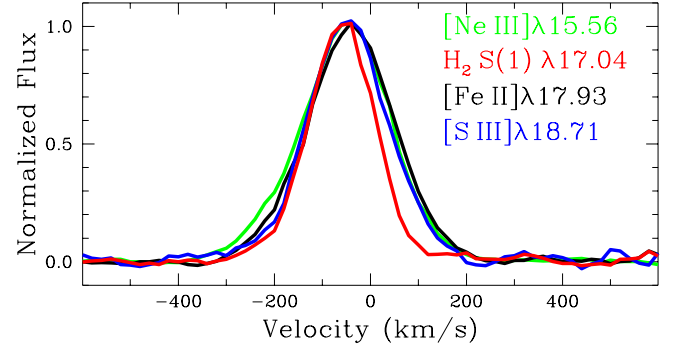}}}
\caption{
Normalized profiles of FIR lines falling at wavelengths where the SWS
spectrometer achieves its highest spectral resolution. The nominal
resolving power for extended sources is 222, 198, 186 and 175 km/s
for the [NeIII], H$_2$, [FeII] and [SIII] lines, respectively.
The ionic lines are all resolved and significantly broader 
than H$_2$ which is narrow
and unresolved. This result agrees with the ground based velocity maps
(Fig.~\ref{velcuts}) and indicate that the ionized lines are produced
downstream of the shock front, while H$_2$ arises from the precursor
(see text, Sect.~\ref{observations}).
}
\label{vel}
\end{figure}

\begin{figure}
\centerline{\resizebox{\hsize}{!}{\rotatebox{0}{\includegraphics{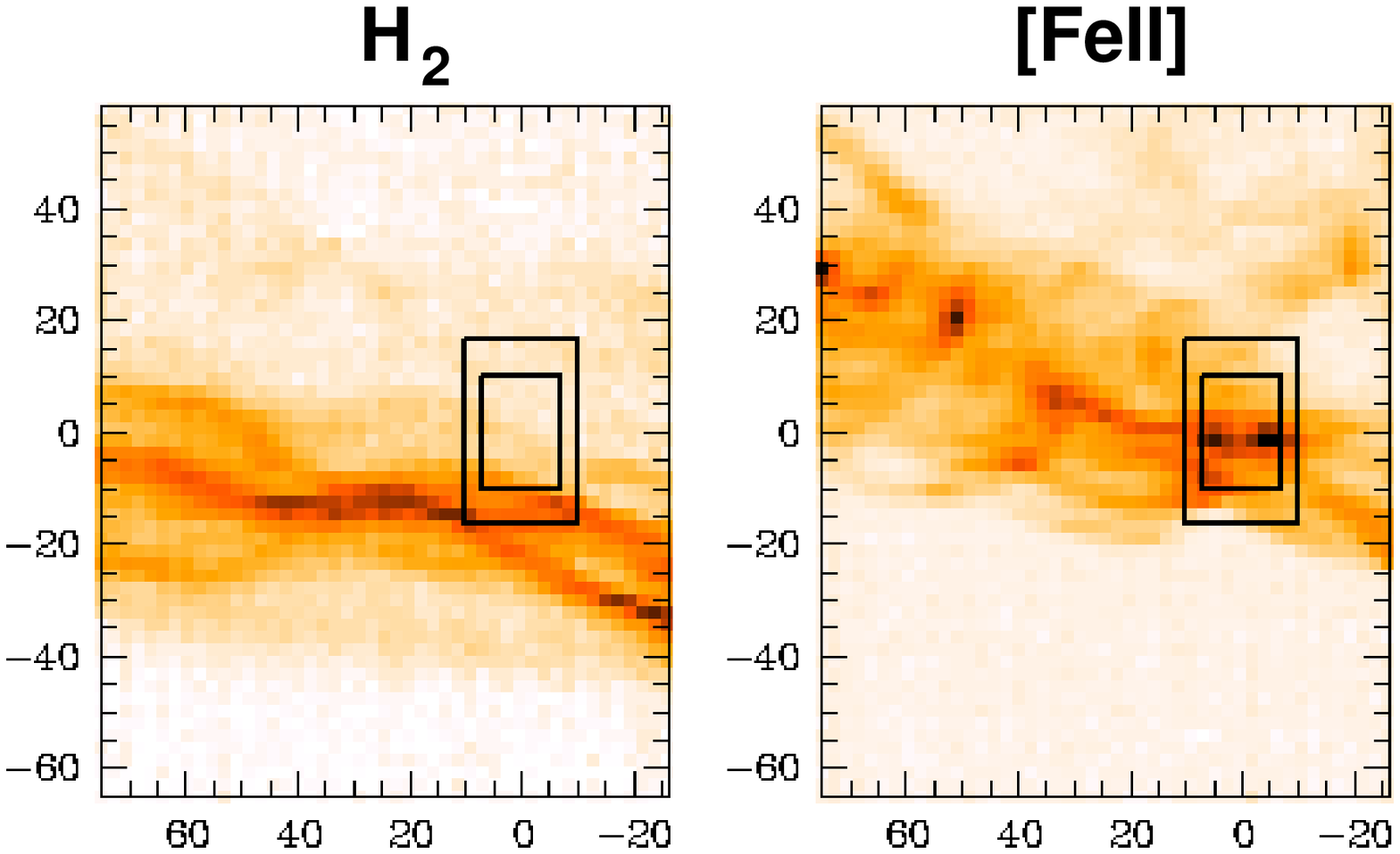}}}}
\caption{
Line images of H$_2$ (2.121 $\mu$m) and [FeII] (1.644 $\mu$m) reconstructed
from the IRSPEC spectra described in Sect.~\ref{observations}.
Coordinates are arcsec from the `ionized peak' 
where all the ISO spectra presented here were centered. 
The sizes of the black rectangles correspond to the smallest 
and largest slits of SWS.
The ground based
[FeII]\L1.644 flux is 70, 82, and 120 ($\times 10^{-20}$ W cm$^{-2}$)
in the 14x20, 14x27 and 20x33 sqarcsec SWS apertures, respectively.
These numbers are used to compute the aperture correction factors
listed in Table~\ref{tab_flux}.
%This figure  is also available at http://www.arcetri.astro.it/$\sim$oliva
}
\label{irspec}
\end{figure}

The final rebinned spectra are displayed in Figs.~\ref{sws02} to \ref{phts}
and the derived line fluxes
are listed in Table~\ref{tab_flux} together with additional information.
Note that most of the lines in the 2.4--40 $\mu$m range
were observed twice, i.e. in the
complete SWS01 spectrum (Fig.~\ref{sws01}) and in the deeper
SWS02 line scans (Fig.~\ref{sws02}). The derived fluxes were 
always within $\pm$30\% and in most cases equal to much better than 20\%.
The errors quoted in Table~\ref{tab_flux} also include the differences
between the two measurements.
It should be noted that, to the best of our knowledge, the 
transitions of [ClII]\L14.36 and
[PII]\L32.8 are newly detected astronomical lines. 
Of interest
is also the marginal detection of an unidentified feature at 74.2 $\mu$m
whose position and flux are remarkably similar to those found in the
spectrum of NGC7027 (Liu et al. \cite{liu}).

The SWS spectrum (Figs.~\ref{sws02}, \ref{sws01}) is characterized by prominent 
lines over a faint
continuum (about 0.5 Jy at 10 $\mu$m)  while emission by
cold dust is evident in the LWS spectrum (Fig.~\ref{lws}).
The level of the 100 $\mu$m
continuum is similar to the background IRAS level reported by Arendt 
(\cite{arendt})
and the continuum seen in the LWS spectrum is, therefore,
 probably dominated by back/foreground emission
from the galaxy disk.
The same applies to the PAH's features visible in the PHT-S
spectrum (Fig.~\ref{phts})
and to the [OI]\L63.2, [NII]\L121 and [CII]\L157 
lines which are likely to be strongly contaminated by emission from
the diffuse ISM.

Given the relatively large extinction towards RCW103,  i.e.
A$_{\rm H}\!\simeq\!0.75$ or A$_{\rm V}\!\simeq\!4.5$ (cf. Oliva et al.
1990, hereafter \cite{OMD90}), the reddening corrections are not 
negligible for the lines at the shortest wavelengths. Therefore, in 
Table~\ref{tab_flux} we also list the 
extinctions values which were derived assuming a `typical' reddening curve,
i.e.  $\tau\!\approx\!\lambda^{-1.75}$ outside of the silicate band.
The largest and most uncertain correction is for [SIV]\L10.52 which lies 
within the silicate band and for which we have assumed 
A(10$\mu$m)/\AV$\simeq$0.1.

Useful dynamical information can be derived from the profiles of the
lines between 15 and 19 $\mu$m, the wavelength
range where SWS02 grating spectra achieve their highest spectral resolution.
The instrumental resolution depends on the size of the source
along the dispersion direction and varies, for example, between 120 and
175 km/s for a compact and extended [SIII]\L18.7 source, respectively.
Luckily, the SWS slit was almost exactly
aligned N--S and was, therefore,  roughly uniformly illuminated in
the dispersion (E--W) direction (cf. Fig.~\ref{irspec}).
The observed line profiles are displayed
in Fig.~\ref{vel}. The ionic lines are resolved and
exhibit similar profiles, within the errors, but are broader
than the H$_2$ line which is unresolved.
This agrees well with the higher spatial resolution velocity
maps from the NIR data discussed below.

\subsection{ Ground based near infrared spectroscopy }

Near infrared observations were collected
in Jan 1992 at the ESO--NTT using the long--slit spectrometer IRSPEC 
equipped with a 62x58
SBRC InSb array whose pixel size was 2.2\arcsec\ along the slit
and $\simeq$5 \AA\  along the dispersion direction.
Line images of [FeII]\L1.644 and H$_2$~\L2.121 
were reconstructed from 57 spectra with the 
100\arcsec~x~2.2\arcsec\ slit aligned E-W and
shifted by steps of 2.2\arcsec\ along the N-S direction.
Each long-slit spectrum consisted of a single on--chip integration of 30 sec
with sky exposures every 10 spectra.

\begin{figure}
\centerline{\resizebox{\hsize}{!}{\rotatebox{0}{\includegraphics{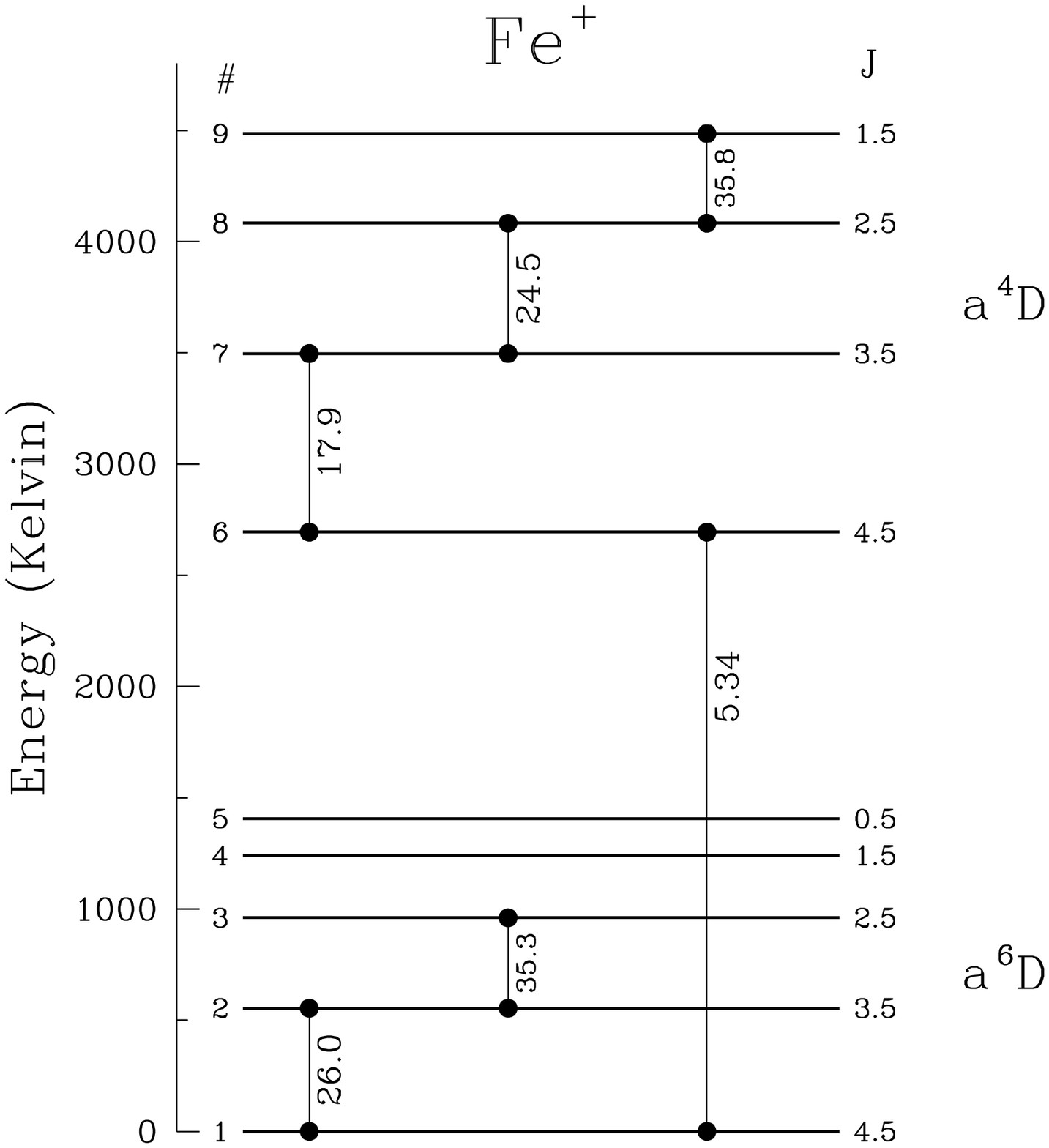}}}}
\caption{ Energy level diagram showing the paths of the
[FeII] lines detected by ISO. The near IR lines come from higher energy levels
which are visualized in Fig.~6b of \cite{OMD90}.
}
\label{grotrian}
\end{figure}

\begin{figure}
\centerline{\resizebox{\hsize}{!}{\rotatebox{0}{\includegraphics{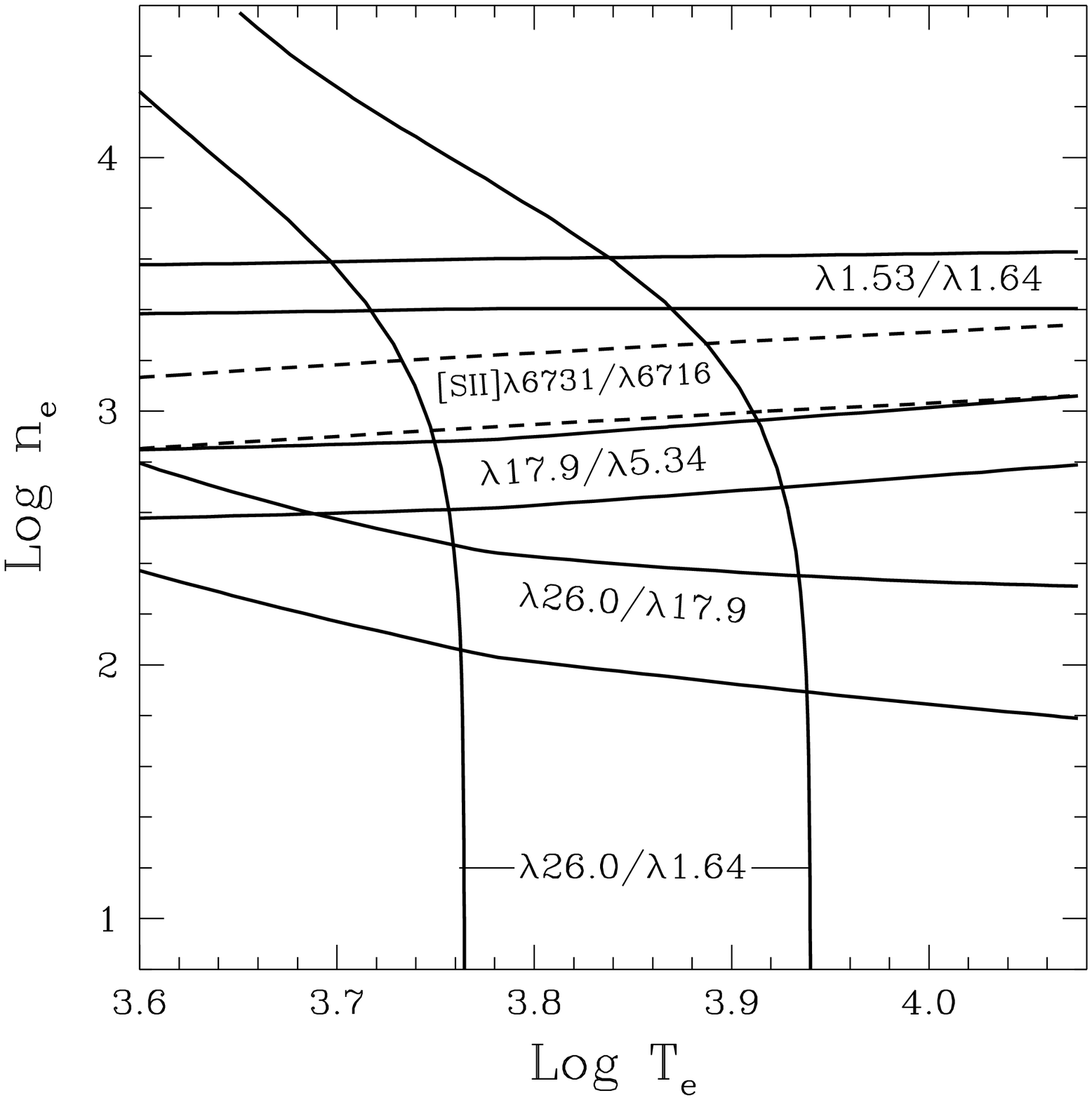}}}}
\caption{
Contour plots of the [SII] and [FeII] diagnostic line ratios listed in 
Table~\ref{tab_diag}. The curves for [FeII] are computed using  the collision
strengths of Pradhan \& Zhang (\cite{pradhan_zhang}) and the transition 
probabilities of Nussbaumer \& Storey (\cite{nussbaumer88})
and Quinet et al. (\cite{quinet}).
}
\label{fe_diag}
\end{figure}
The integrated line images are displayed in Fig.~\ref{irspec} whose
caption also include results of [FeII] aperture photometry which has been
used to determine the 
correction factors for the different beams used by SWS (see
also Table~\ref{tab_flux}). 
These assume that all ionized species have spatial distribution similar to
[FeII] which is justified by the following arguments.\\
-- The flux of \BA\  seen by ISO is within $<$10\% of the value extrapolated
from the ground based measurement (\cite{OMD90}) assuming 
a constant [FeII]/\BA\ ratio
over the region of interest.\\
-- The morphology of the [FeII] filaments is
virtually identical to those seen in optical line images
(cf. e.g. Moorwood et al. \cite{moorwood87}).\\
It should be noted, however, that the H$_2$ lines arise from a totally
different region $\sim$20\arcsec\  south of the ionized gas and
outside the optical/radio/X--ray remnant,
as originally found by \cite{OMD89}. This indicates that H$_2$ emission
traces material which has not yet been reached
by the shock, most probably a molecular cloud heated by the soft
X--rays from the shock front (cf. \cite{OMD90}).

Fig.~\ref{velcuts} shows images at various velocity bins, each roughly
corresponding to the wavelength range covered by 1 pixel. Evident
are the high velocity [FeII] filaments whose projected
velocities extend up to $\pm$250 km/s and are compatible with
the idea that this line is produced downstream of the fast shock 
(cf. the Introduction).
The H$_2$ filaments, on the contrary, do not show evidence of motions
larger than the FWHM=130 km/s instrumental resolution.

\section{ Physical parameters of the emitting gas }

\label{physical_param}

\subsection{ Temperature and density}
\label{temperature}

\begin{table}
\caption{Diagnostic line ratios}
\label{tab_diag}
\def\SKIP#1{\noalign{\vskip#1pt}}
\def\UNC{\rlap{:}}
\def\UNO{\rlap{$^{(1)}$}}
\def\DUE{\rlap{$^{(2)}$}}
\def\TRE{\rlap{$^{(3)}$}}
\def\NA{\rlap{$^a$}}
\def\NB{\rlap{$^b$}}
\def\NC{\rlap{$^c$}}
\def\ND{\rlap{$^d$}}
\def\NE{\rlap{$^e$}}
\def\X{ $\times$ }
\def\SB{$\rm [$}
\begin{flushleft}
\begin{tabular}{lrcc}
\hline\hline
\SKIP{1}
 Line ratio & Observed\UNO &   \Ne\DUE & \Te\DUE  \\
            &              &  (cm$^{-3}$) &  (x$10^3$ K) \\
\SKIP{2}
\hline
\SKIP{1}
[OIII]\L51.7/\L88.2 & $-.07\pm.09$  & 30--250  & -- \\
\SKIP{0}
[NeIII]\L15.6/\L36.0 & $1.0\pm.08$  & $\le$5000  & -- \\
\SKIP{0}
%
% [NeIII]\L3869/Hbeta=0.41 (LD), [NeIII]\L15.6/Bralpha=13, Hbeta/Bralpha=12.9
%        =>[NeIII]\L3869/\L15.6=0.41
%  Remember to say that Dopita predict T(NeIII)~8500 at vs>=300 km/s
%
[NeIII]\L3869/\L15.6\NA & $-.40\pm.20$ & -- & 8.3--10.5 \\
\SKIP{0}
[NeV]\L14.3/\L24.3 & $-.15\pm.13$  & $\le$5000 & -- \\
\SKIP{0}
[SIII]\L18.7/\L33.5 & $-.05\pm.08$ & $\le$1200 & -- \\
%
% [SIII]\L9532/Hbeta=0.16 (Dennefeld), [SIII]\L18.7/Bralpha=4.25,
%             Hbeta/Bralpha=12.9
%        =>[SIII]\L9532/\L18.7=0.49
%  Remember to say that Dopita predict T(SIII)~?? at vs>=300 km/s
%
\SKIP{0}
[SIII]\L9531\L18.7\NA & $-.31\pm.20$ & -- & 7.1--13.8 \\
\SKIP{0}
[SII]\L6731/\L6716\NB & $.11\pm.05$ & 800--2000 & -- \\
\SKIP{0}
[FeII]\L35.3/\L26.0 & $-.56\pm.09$ &  any & -- \\
\SKIP{0}
[FeII]\L24.5/\L17.9 & $-.60\pm.11$ &  any & -- \\
\SKIP{0}
[FeII]\L17.9/\L5.34 & $-.16\pm.13$ & 400--1000 & -- \\
\SKIP{0}
[FeII]\L17.9/\L26.0 & $-.24\pm.07$ & 80--320 & \\
\SKIP{0}
[FeII]\L26.0/\L1.64\NC & $-.26\pm.13$ & -- & 5.2--8.7 \\
\SKIP{0}
[FeII]\L1.53/\L1.64\ND & $-.92\pm.06$ & 2500--4100 & -- \\
\SKIP{2}
\hline
\SKIP{2}
\end{tabular}
\def\NOTA#1#2{
\hbox{\vbox{\hbox{\hsize=0.030\hsize\vtop{\centerline{#1}}}}
      \vbox{\hbox{\hsize=0.97\hsize\vtop{\baselineskip4pt #2}}}}\vskip1pt}
\NOTA{ $^{(1)}$ }{ Log of observed ratio,
including aperture and reddening corrections
(cf. Table~\ref{tab_flux} and Sect.~\ref{observations}) }
\NOTA{ $^{(2)}$ }{ Deduced gas density and temperature, see also 
Fig.~\ref{fe_diag} }
\NOTA{ $^a$ }{ From [NeIII]\L3869/\HB=0.41 (Leibowitz \& Danziger
\cite{leibowitz}) and  [SIII]\L9531/\HB=0.16 (Dennefeld \cite{dennefeld})
adopting the case--B \HB/\BA=13 ratio}
\NOTA{ $^b$ }{ [SII] lines from Leibowitz \& Danziger (\cite{leibowitz}) }
\NOTA{ $^c$ }{ Flux of [FeII]1.644 from IRSPEC image (cf. caption of
Fig.~\ref{irspec}) corrected for extinction using A$_{\rm H}$=0.75 }
\NOTA{ $^d$ }{ Near infrared [FeII] lines from \cite{OMD90} }
\end{flushleft}
\end{table}
The ISO spectra include several density and/or temperature
sensitive line pairs.
Besides the well known, \Ne--sensitive [OIII], [NeIII], [NeV] and [SIII]
fine structure lines, useful transitions are those of [FeII] whose paths are
displayed in Fig.~\ref{grotrian}, while their dependence on \Te\
and \Ne\  is visualized in Fig.~\ref{fe_diag}. Note that
the ratios [FeII]\L35.3/\L26.0 and [FeII]\L24.5/\L17.9 are also 
density sensitive,
but vary by only a factor $\simeq$1.7 between the low and the high density
limits so the resulting \Ne\ is highly  uncertain.

The deduced values of \Te\ and \Ne\  are summarized in 
Table~\ref{tab_diag} where the most remarkable result is the large
spread of densities deduced from the various diagnostic ratios.
The lowest value is from the [OIII] lines which are measured through 
the much larger LWS beam (cf. Table~\ref{tab_flux}) and are probably
affected by the emission of a diffuse, lower density component.
Some of the discrepancies between the [FeII] densities could also
reflect density and temperature stratification. In particular,
the ground state line at 26.0 $\mu$m could partially arise from excitation
by collisions with atomic hydrogen in regions
far downstream of the shock where the temperature is too low
to produce the other [FeII] lines. However, such a cold component
should primarily show up in the temperature sensitive
[FeII]\L26.0/\L1.64 ratio which, on the contrary, yields a relatively
high value for \Te\  and therefore implies
that any cold component can account, at most, for 30\% of the
observed flux of [FeII]\L26.0.
An alternative explanation is that the different [FeII] densities simply
reflect relatively small
uncertainties in the atomic parameters of Fe$^+$ which, for example,
overestimate the critical density of [FeII]\L1.53 and of the other
satellite lines in the near infrared.

It should be noted that much larger discrepancies between the [FeII]
densities from FIR and NIR line ratios were found from observations
of the Galactic center (GC, Lutz et al.~\cite{galactic_center})
where, however, all the [FeII] lines but one are reasonably well
explained by a collisionally excited plasma
with \Te$\sim$8000 K and \Ne=$10^{3-4}$ cm$^{-3}$. The only
line which does not fit is 
%[FeII]\L17.9 
\L17.9 
which is a factor of $\simeq$10
lower than predicted.
This cannot be easily explained neither by alternative
excitation mechanisms such as e.g. UV pumping (Lucy private communication)
nor by advocating large uncertainties on the collision strengths which
are ruled out by the results presented here. 
A partial explanation to the ``GC anomaly''  could be to assume that the 
foreground extinction at 17.9 $\mu$m is a factor of 2--3 larger than that
predicted from the standard reddening curve. \\
%This could either be due
%to an underestimated cross section of the 18$\mu$m silicates band 
%or to absorption by molecular or ice bands.  \\

The [NeIII], [NeV] and [SIII] ground 
state lines ratios 
are compatible, within the errors, with their low density limits, but
useful estimates of temperature can be obtained including
measurements of optical transitions from the same ions.
The values listed in Table~\ref{tab_diag} are derived from the available
fluxes of optical and FIR lines relative to \HB\ and \BA, respectively,
and assume a generous error of $\pm$0.2 dex to include e.g. possible
uncertainties in the reddening correction of the optical data.
The resulting temperatures are quite well constrained, nevertheless, and
the values are compatible with those predicted for the post--shock
region of shocks with velocities $\ge$300 km/s
(cf. Tables 3A--3D of \cite{DS96}).
Note in particular that slower shocks are predicted to have a post--shock
[NeIII]
temperature of 15--30 $\times10^3$ K and significantly hotter than the
observed value.

The temperature of [NeV] is of particular interest to verify if this ion
is produced downstream of the shock, in which case NeV is predicted to 
be collisionally ionized in plasma at about $3.5\,10^5$ K, or in the
much cooler ($<\!2\,10^4$ K) photoionized precursor. In the first case
one expects [NeV]\L3426/\L14.3=8 or, equivalently, [NeV]\L3426/\HB=0.1,
while values a factor of 6--7 lower are expected if the emission is from
the precursor.
Unfortunately,
no observation of [NeV]\L3426 is available in the literature and, given
the quite high extinction, measurements at the required depth (a few \%
of \HB) are not straightforward.

\subsection{ Abundances }
\label{abundances}

\begin{table}
\caption{Relative abundances}
\label{tab_abun}
\def\MYBOX#1#2#3{\hbox to 75pt{#1 \hfil #2 \hfil #3}}
\def\MYbox#1#2#3{\hbox to 60pt{#1 \hfil #2 \hfil #3}}
\def\SKIP#1{\noalign{\vskip#1pt}}
\def\UNC{\rlap{:}}
\def\UNO{\rlap{$^{(1)}$}}
\def\DUE{\rlap{$^{(2)}$}}
\def\TRE{\rlap{$^{(3)}$}}
\def\NA{\rlap{$^a$}}
\def\NB{\rlap{$^b$}}
\def\NC{\rlap{$^c$}}
\def\X{ $\times$ }
\begin{flushleft}
\begin{tabular}{lll}
\hline\hline
\SKIP{1}
 \hfil Lines used\UNO\hfil  &  \MYBOX{}{Abundance ratio\DUE}{}  &  
		\MYbox{}{Solar value}{} \\
\SKIP{2}
\hline
\SKIP{1}
%
% lg(FeIII22.9/FeII26.0)=-1.05+/-0.07
% lt=3.72 lj(26.0)=-20.13 lj(22.9)=-20.26 => log(Fe++/Fe+) = -0.92
% lt=3.94 lj(26.0)=-20.22 lj(22.9)=-20.28 => log(Fe++/Fe+) = -1.05
%
[FeIII]22.9/[FeII]26.0 & \MYBOX{[Fe$^{++}$/Fe$^+$]}{=}{$-0.99$} & 
			\MYbox{}{--}{} \\

\SKIP{0}
%
% lg(NeIII15.6/NeII12.8)=-0.05+/-0.07
% lt=3.72 lj(15.6)=-20.46 lj(12.8)=-20.96 => log(Ne++/Ne+) = -0.55
% lt=3.94 lj(15.6)=-20.52 lj(12.8)=-21.02 => log(Ne++/Ne+) = -0.55
% extreme: if lt(NeIII)=4.2 lj(15.6)=-20.62 and lt(NeII)=3.7 lj(12.8)=-20.95
%                                         => log(Ne++/Ne+) = -0.38
%
[NeIII]15.6/[NeII]12.8 & \MYBOX{[Ne$^{++}$/Ne$^+$]}{=}{$-0.47$} & 
			\MYbox{}{--}{} \\
\SKIP{0}
%
% lg(NeV14.3/NeIII15.6)=-1.92+/-0.23
% lt=4.2  lj(14.3)=-19.92 lj(15.6)=-20.62 => log(NeV-cold/NeIII) = -2.62
% lt=5.5  lj(14.3)=-20.96                 => log(NeV-cold/NeIII) = -1.58
% extreme: if lt(NeIII)=4.2 lj(15.6)=-20.62 and lt(NeII)=3.7 lj(12.8)=-20.95
%                                         => log(Ne++/Ne+) = -0.38
%
%[NeV]14.3/[NeIII]15.6 & \MYBOX{[Ne$^{+4}$/Ne$^{++}$]}{=}{$-2.6$} & 
%			\MYbox{}{--}{} \\

\SKIP{0}
%
% lg(ArIII8.99/ArII6.98)=-0.95+/-0.08
% lt=3.72 lj(8.99)=-19.98 lj(6.98)=-19.81 => log(Ar++/Ar+) = -0.78
% lt=3.94 lj(8.99)=-20.02 lj(6.98)=-19.83 => log(Ar++/Ar+) = -0.76
% extreme: if lt(ArIII)=4.2 lj(8.99)=-20.10 and lt(ArII)=3.7 lj(6.98)=-19.81
%                                         => log(Ar++/Ar+) = -0.66
%
[ArIII]8.99/[ArII]6.98 & \MYBOX{[Ar$^{++}$/Ar$^+$]}{=}{$-0.72$} & 
			\MYbox{}{--}{} \\

\SKIP{0}
%
% lg(SIV10.5/SIII18.7)=-0.95+/-0.11
% lt=3.72 lj(10.5)=-19.22 lj(18.7)=-19.93 => log(S+3/S++) = -1.66
% lt=3.94 lj(10.5)=-19.26 lj(18.7)=-19.96 => log(S+3/S++) = -1.65
% extreme: if lt(SIV)=4.2 lj(10.5)=-19.52 and lt(SIII)=3.7 lj(18.7)=-19.93
%                                         => log(S++/S+) = -1.36
%
[SIV]10.5/[SIII]18.7 & \MYBOX{[S$^{+3}$/S$^{++}$]}{=}{$-1.5$\UNC} & 
			\MYbox{}{--}{} \\

\SKIP{4}
%
% lg(NIII57.3/OIII51.6)=-0.52+/-0.12
% j(51.6)/j(57.3) depns on density, =0.8 below ne=100 and =1.8 above ne=1e4
% take 0.8 because we should be dominated by the diffuse component.
%                              => log(N++/O++)=-0.62
[NIII]57.3/[OIII]51.7  & \MYBOX{[N$^{++}$/O$^{++}$]}{=}{$-0.6$\UNC} & 
				\MYbox{[N/O]}{=}{$-0.89$} \\

\SKIP{4}
%
% lg(FeII17.9/Bra)=0.79+/-0.08
% lg(FeII17.9/Bra)=0.79+/-0.08
% lt=3.72 lj(17.9)=-20.76 lj(Bra)=-25.70 => 12+log(Fe+/H+) = 7.85
% lt=3.94 lj(17.9)=-20.72 lj(Bra)=-25.94 => 12+log(Fe+/H+) = 7.57
%
%
% Serve veramente questo? O e' forse meglio NeII/HII?
%
%[FeII]17.9/Br$\alpha$   & [Fe$^+$/H$^+$)=$-4.3\pm0.2$ &
%             Fe/H=$-4.49$ \\
% lg(NeII12.8/Bralpha)=1.16+/-0.07
% lt=3.72 lj(12.8)=-20.96 lj(Bra)=-25.70
% lt=3.94 lj(12.8)=-21.02 lj(Bra)=-25.94
%                                         => log(Ne+/H+)=-3.82 - -3.52
[NeII]12.8/Br$\alpha$  & \MYBOX{[Ne$^+$/H$^+$]}{=}{$-3.7$\UNC} & 
				\MYbox{[Ne/H]}{=}{$-3.93$} \\

\SKIP{0}
%
% lg(FeII26.0/SiII34.8)=-0.33+/-0.07
% lt=3.72 lj(26.0)=-20.13 lj(34.8)=-19.75 => log(Fe+/Si+) = 0.05
% lt=3.94 lj(26.0)=-20.22 lj(34.8)=-19.85 => log(Fe+/Si+) = 0.04
%
%\SKIP{0}
%[FeII]26.0/[SiII]34.8 & [Fe$^+$/Si$^+$)=$0.04\pm0.09$ & 
%				[Fe/Si)=$-0.04$ \\
[SiII]34.8/[NeII]12.8  & \MYBOX{[Si$^+$/Ne$^+$]}{=}{$-0.66$} & 
				\MYbox{[Si/Ne]}{=}{$-0.51$} \\

\SKIP{0}
%
%
% lg(PII32.8/NeII12.8)=-1.94+/-0.10
% lt=3.72 lj(32.8)=-20.32 lj(12.8)=-20.96
% lt=3.94 lj(32.8)=-20.37 lj(12.8)=-21.02
%                                         => log(P+/Ne+)=-2.64 - -2.53
%
[PII]32.8/[NeII]12.8 & \MYBOX{[P$^+$/Ne$^+$]}{=}{$-2.6$\UNC} &
					\MYbox{[P/Ne]}{=}{$-2.6$} \\

\SKIP{0}
%
% lg(SIII18.7/NeIII15.6)=-0.51+/-0.07
% lt=3.72 lj(18.7)=-19.93 lj(15.6)=-20.46
% lt=3.94 lj(18.7)=-19.96 lj(15.6)=-20.52
%                                         => log(S++/Ne++)=-1.1 - -1.01 
%
[SIII]18.7/[NeIII]15.6 & \MYBOX{[S$^{++}$/Ne$^{++}$]}{=}{$-1.0$} &
					\MYbox{[S/Ne]}{=}{$-0.86$} \\

\SKIP{0}
%
%
% lg(ClII14.4/NeII12.8)=-1.85+/-0.10
% lt=3.72 lj(14.4)=-20.15 lj(12.8)=-20.96
% lt=3.94 lj(14.4)=-20.22 lj(12.8)=-21.02
%                                         => log(Cl+/Ne+)=-2.72 - -2.59
%
[ClII]14.4/[NeII]12.8 & \MYBOX{[Cl$^+$/Ne$^+$]}{=}{$-2.59$} &
					\MYbox{[Cl/Ne]}{=}{$-2.5$} \\

\SKIP{0}
%
% lg(ArII6.98/NeII12.8)=-0.22+/-0.08
% lt=3.72 lj(6.98)=-19.81 lj(12.8)=-20.96  
% lt=3.94 lj(6.98)=-19.83 lj(12.8)=-21.02 
%                                         => log(Ar+/Ne+)=-1.43 - -1.35
%
[ArII]6.98/[NeII]12.8 & \MYBOX{[Ar$^+$/Ne$^+$]}{=}{$-1.39$} &
				\MYbox{[Ar/Ne]}{=}{$-1.47$} \\

\SKIP{0}
%
% lg(FeII17.9/NeII12.8)=-0.37+/-0.07
% lt=3.72 lj(17.9)=-20.76 lj(12.8)=-20.96 
% lt=3.94 lj(17.9)=-20.72 lj(12.8)=-21.02
%                                         => log(Fe+/Ne+)=-0.67 - -0.57
%
[FeII]17.9/[NeII]12.8 & \MYBOX{[Fe$^+$/Ne$^+$]}{=}{$-0.62$} & 
				\MYbox{[Fe/Ne]}{=}{$-0.55$} \\

%\SKIP{0}
%
% Could be better to compare it with NeII which has a much higher 
% critical density
%
% lg(NiII6.64/FeII17.9)=-0.85+/-0.10
% lt=3.72 lj(6.64)=-21.22 lj(17.9)=-20.76 
% lt=3.94 lj(6.64)=-21.23 lj(17.9)=-20.72
%                                         => log(Ni+/Fe+)=-0.39 - -0.34
%
%[NiII]6.64/[FeII]17.9 & [Ni$^+$/Fe$^+$]=$-0.36$ &
%					[Ni/Fe]=$-1.27$ \\

\SKIP{0}
%
% Remember that n_crit(NeII)=6.6e5 and n_crit(NiII)=7.9e6
%
% lg(NiII6.64/NeII12.8)=-1.23+/-0.10
% lt=3.72 lj(6.64)=-21.22 lj(12.8)=-20.96
% lt=3.94 lj(6.64)=-21.23 lj(12.8)=-21.02
%                                         => log(Ni+/Ne+)=-1.03 - -0.96
%
[NiII]6.64/[NeII]12.8 & \MYBOX{[Ni$^+$/Ne$^+$]}{=}{$-1.03$} &
				\MYbox{[Ni/Ne]}{=}{$-1.82$} \\

\SKIP{1}
\hline
\SKIP{2}
\end{tabular}
\def\NOTA#1#2{
\hbox{\vbox{\hbox{\hsize=0.030\hsize\vtop{\centerline{#1}}}}
      \vbox{\hbox{\hsize=0.97\hsize\vtop{\baselineskip2pt #2}}}}\vskip2pt}
\NOTA{ $^{(1)}$ }{ Fluxes from Table~\ref{tab_flux} including
aperture and reddening corrections }
\NOTA{ $^{(2)}$ }{ Logaritmic values, typical uncertainties are $\pm0.12$
dex except for those marked with a column which are $\pm0.25$ dex }
\end{flushleft}
\end{table}
A specific advantage of FIR lines is that their emissivities depend 
very little on the gas temperature and abundances derived from
these lines are, therefore, little affected by
uncertainties in the assumed value of \Te.
%The only exception is NeV whose relative abundance depends on whether...
%(serve metterlo?)
%The values listed in Table~\ref{tab_abun} are derived assuming 
%5000$\le$\Te$<$10000 for the singly ionized species while
%7000$\le$\Te$\le$15000
%for double ionized 
Also, to minimize the effect of density
variations, we selected for each ion the line
with the largest critical density which, for most species, was $\gg10^3$
cm$^{-3}$ and larger than the values of \Ne\  derived above.
The derived abundances are listed in Table~\ref{tab_abun} where the quoted
errors also include generous uncertainties on temperature
(i.e.  \Te=5000--10000 K and \Te=7000--15000 K for singly and double ionized
species, respectively) and density ($\Ne\!=\!0-4000$ cm$^{-3}$).

The most remarkable result is that, apart
for the anomaly of nickel which is discussed below,
all the metal abundances are close to the solar
values.
This indicates that the emitting gas is indeed ISM material in which
the dust grains have been destroyed by the shock front thus returning
all the refractory species (Si, Fe) into the gas phase.
It should also be noted that, to the best of our knowledge, the ISO data 
provide for the first time a measurement of the abundances of 
phosphorus and chlorine in any supernova remnant. 
Moreover, the 
abundance of silicon from the FIR [SiII] line is much less uncertain than those
normally derived from UV lines whose intensities strongly depend on the
assumed gas temperature (e.g. Russel \& Dopita \cite{russel}).

\subsubsection{ The nickel anomaly }

The relatively large strength of the optical [NiII]\L7379 line
in SNRs and other nebulae
has long puzzled astronomers because it requires a Ni$^+$/Fe$^+$
abundance ratio a factor $\sim$10 larger than the cosmic value.
A similar overabundance, i.e. [Ni$^+$/Fe$^+$]=\-$-0.41$ dex or
a factor of 7 above the solar [Ni/Fe]=$-1.27$ value, is inferred here from
the [NiII]\L6.64 ground state transition  
(cf. Table~\ref{tab_abun}).
This result is very difficult to understand because the line emitting
gas is shock excited interstellar material which should not have,
therefore, an anomalous Ni abundance.
 A number of explanations for the optical result have been
proposed (cf. Bautista et al. \cite{bautista} for a recent review).
These include:\\
-- Selective dust depletion of Fe relative to Ni, which cannot
hold in our case because most of the iron is in the gas phase. \\
-- Low Fe$^+$/Fe relative abundances, which can be excluded
because most of the iron is in the form of Fe$^+$ 
(cf. Table~\ref{tab_abun}). \\
-- Contribution by UV fluorescence to the [NiII] lines, which could be 
important
for the optical lines but has negligible effect on
the [NiII]\L6.64 ground state transition
(cf. Table 1 of Lucy \cite{lucy}). Also, no obvious source of strong
UV radiation exists in RCW103 (cf. Sect.~4.1.3  of \cite{OMD90}).
\\
-- Density stratification with low \Ne\ regions emitting [SII], higher
density gas emitting [FeII] and the densest ($\sim$10$^6$ cm$^{-3}$) regions
dominating [NiII] whose lines have the highest critical density
(Bautista et al. \cite{bautista}). 
However, such a high density component should also (but does not) show up in 
other lines with high critical density, in particular [NeII]\L12.8
which has $n_{crit}\!=\!6\,10^5$ cm$^{-3}$ and only a factor of 10
lower than the critical density of the [NiII]\L6.64 line. Since 
the [NiII]/[NeII] ratio also requires a nickel overabundance of a factor of 6
(cf. Table~\ref{tab_abun}), and considering also the relatively low
densities derived from the [FeII] lines (cf. Table~\ref{tab_diag}),
one is  forced to assume that most
of the nickel is in regions with densities $\ga\!10^7$ cm$^{-3}$ while
all the other species, including Fe, must be in regions with densities
$<\!10^4$ cm$^{-3}$, which is implausible.

We therefore conclude that the apparently large Ni$^+$/Fe$^+$ relative
abundance simply reflects uncertainties in the collision strengths for
[NiII] whose computed values are systematically a factor $\approx$10
lower than those of [FeII].  In particular, the collision strength of the
[NiII]\L6.64 ground state line is only $\Upsilon$=0.15 
and by far (i.e. a factor of $>$5) lower than the collision strengths of
the main ground state transitions of any other astrophysical 
abundant species.
Adopting, for example, $\Upsilon$=1.0 for the [NiII]\L6.64 transition
would yield [Ni$^+$/Ne$^+$]=$-1.85$ and [Ni$^+$/Fe$^+$]=$-1.18$, i.e. 
values within 0.1 dex of the solar Ni/Ne and Ni/Fe relative abundances.

\section{ Discussion }
\label{discussion}

\subsection{ Comparison with shock models }
\label{shock_models}

Explicit predictions for a few FIR
lines have been included in the models for relatively slow ($\le$150 km/s)
%
% Modified
%
shock interacting with low density material by
Raymond (\cite{raymond}) and Shull \& McKee 
(\cite{shull_mckee}). The first lists [NeII], [SiII] and [FeII]
while the latter include [SiII], [SIII] and [SIV].
The predicted line ratios from the above mentioned models are not in good
agreement with our results.
In particular, the observed [SIV]/[SIII] ratio is a factor $>$5 
larger than the computed values.
Also, most models predict a factor of $>$3
too strong [SiII] (relative to [NeII], [FeII], [SIII]), but this
could be attributed to uncertainties in the atomic parameters of SiII
which have been updated several times since the publication of
the shock model results.
The most important discrepancy, however, is that the predicted 
surface brightnesses are always
a factor $>$10 lower than the observed values, which simply reflects the
fact that the shock of RCW103 is much faster than
the values used in the above models (cf. the Introduction and below).

%
% Modified
%
Models of slow shocks interacting with very dense gas 
($n\!\ge\!10^3$ cm$^{-3}$) were developed by Hollenbach \& McKee 
(\cite{hollenbach89}) who also include explicit predictions for all
the FIR lines of singly ionized and neutral species. The main problem
with these models is that they span preshock densities much larger than 
the $<$300 cm$^{-3}$ required to account for the measured electron densities
in the post--shock region. Consequently, lines with low critical densities,
e.g. [SiII], are predicted too faint. Moreover, the models underestimate the
flux of [FeII] lines by a factor of $\sim$5.\\

The most recent models of \cite{DS96}, which cover shock velocities of 200--500
km/s and are more representative of the conditions of RCW103, do not
however give explicit predictions for the FIR lines.
Nevertheless, reasonably accurate line ratios can be
computed from the values of ionic column densities and mean temperatures
listed in the above paper.
To a first approximation,
the ratio of two lines from the post--shock region is
$$ {I_1\over I_2} = { N_1\; j_1(T_1) \over N_2\; j_2(T_2) } \eqno{(1)} $$
where $N$ and $T$ are the column densities and temperature of the emitting
ions and $j$ is the line emission coefficient which, for FIR lines,
is very little dependent on the gas temperature. 
The contribution from the photoionized precursor can 
be also computed from the published tables of $N$ and $T$ in the 
pre--shock region, while the compression factor (i.e. the ratio between
the electron density in the post and pre shock regions) can be estimated
imposing that the regions have similar \HB\  surface brightnesses
(cf. Sect. 3.2 of \cite{DS96}), i.e.
$$ \left[\Ne\, N_{p}\right]_{precursor} = 
        \left[\Ne\, N_{p}\right]_{post-shock} \eqno{(2)} $$

\begin{figure}
\centerline{
 \resizebox{\hsize}{!}{\rotatebox{0}{\includegraphics{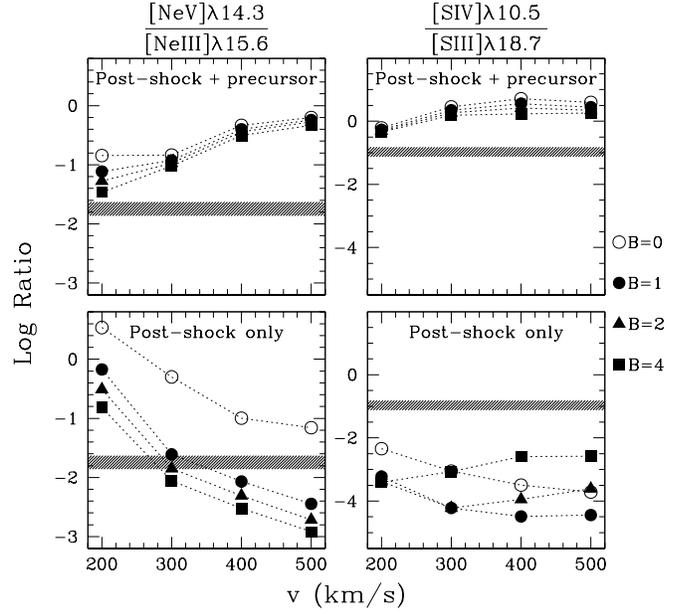}}}}
\caption{ Behaviour of velocity and precursor sensitive line ratios.
The theoretical values are computed from the shock models of \cite{DS96}
as described in Sect.~\ref{shock_models}. B=$B/\sqrt{n}$ is the magnetic 
parameter (cf. \cite{DS96}) and the dashed regions show the observed
values.
}
\label{shock_pred}
\end{figure}
Out of the many FIR line ratios we have identified those which are
most sensitive to the shock speed and to the presence of the photoionized
precursor. 
The behaviour of the selected line ratios
is plotted in Fig.~\ref{shock_pred} where the most remarkable
result is that models including the emission from the precursor
largely overpredict the strength of [SIV] and [NeV]. 
The velocity dependence of the [NeV]/[NeIII] ratio in
the post--shock region may appear at first sight surprising, but can be easily
understood as follows. The post--shock [NeV] emission always occurs
in collisionally ionized gas at $T\!\simeq\!3.5\,10^5$ K and whose
column density, which primarily depends on the shape of the gas
cooling curve, does not strongly vary with the shock speed.
The [NeIII] line, on the contrary, could be strongly enhanced by emission
from photoionized gas in the post--shock region, but this
only occurs at $v\!\ga\!300$ km/s while slower shocks do not produce
enough ionizing photons to support a large NeIII zone. In short,
the sharp decrease of [NeV]/[NeIII] between 200 and 300 km/s is
because the [NeIII] line emission rapidly increases in this velocity
interval.\\

The main conclusion of this analysis is that all ionic lines 
 can be reasonably well reproduced by post--shock emission.
This conclusion also agrees with the broad line profiles observed by SWS
(Sect.~\ref{iso_obs}) and imaging--spectroscopy observations
of [OIII]\L5007 which show complex dynamical structures, similar
to those seen in [FeII] (Fig.~\ref{velcuts}) and incompatible with
emission from the precursor (Moorwood et al. \cite{moorwood87}).
The only ionic line which cannot be accounted for by 
post--shock emission is [SIV] which should be a factor of $>$10 fainter,
but can be reproduced adding 
a quite `incomplete precursor', i.e. $\simeq$5\% the SIV column density of the
precursor predicted by the \cite{DS96} models (cf. Fig.~\ref{shock_pred}).

Given the importance that the photoionized precursor may have in 
modelling the spectra of active galaxy nuclei (Sect.~\ref{agn}),
it is of interest to investigate why little or no ionic
line emission is observed from the precursor of RCW103.
We envisage the following possibilities.\\
-- The precursor in RCW103 is very thin to UV ionizing photons, 
but this is very difficult
to reconcile with the fact that
H$_2$ emission is observed from pre--shock molecular gas
lying outside of the shock front (cf. Fig.~\ref{irspec}).
For H$_2$ to exist, the molecules must be shielded from the
strong field of UV ionizing radiation from the shock front or,
equivalently, the preshock region must be optically thick.\\
%
% Modified
%
-- The shock front in RCW103 is significantly slower than so far
assumed and below $\simeq$150 km/s, the minimum speed required to
produce a prominent phototoionized precursor.
This is in strict constrast with the observed line widths and filament
dynamics (cf. Sect.~\ref{observations} and Fig.~\ref{velcuts}).
Moreover, slow shocks cannot explain the very large
surface brightness of the lines which require a large mechanical
power of the shock, i.e. a large $n\,v_s^3$ product, $n$ being
the preshock density and $v_s$ the shock speed.
More specifically, the average surface brightness of \BA\  within the ISO
beam corresponds to
$\Sigma(\HB)$=$4\, 10^{-3}$ erg cm$^{-2}$ s$^{-1}$ sr$^{-1}$
which, coupled to the predicted values from shock models
(Eq. 3.4 of \cite{DS96}), yields
%
% Modified
%
$$  \left(n\over 100\ {\rm cm^{-3}}\right)
\left(v_s\over 100\ {\rm km/s}\right)^{2.4} \simeq 68\,\cos\theta \eqno{(3)} $$
or, equivalently, a shock speed of about $370\,(\cos\theta)^{0.42}$ km/s 
for a pre--shock
density of 300 cm$^{-3}$. Larger pre--shock
densities are effectively excluded by the measured electron
densities (Table~\ref{tab_diag})
in the post--shock region, i.e. after the gas has been compressed
by the shock front.
The factor $\cos\theta$ takes into account projection effects such as those 
modelled in details by Hester (\cite{hester}) who interpreted the bright 
filaments in IC443 and Cygnus--Loop in therms of relatively slow shocks
seen quasi edge--on and found that small filaments 
amplified by a factor 10--100 should be quite common. However,
this model cannot hold for RCW103 for the following reasons.
This remnant is much brighter (factor of $>$10) than IC443 and Cygnus--Loop.
The {\it average} surface brightness  within the relatively large 
ISO--SWS beam (i.e. the value used in Eq.~3) is already
a factor of $\simeq$4 lower than that observed
on arcsec scales in optical/IR line images of RCW103. 
The most largely amplified edge--on filaments should have  small
radial velocities (FWHM$\la$40 km, cf. Fig.~2 of Hester \cite{hester})
amd this is not compatible with the observed line widths and dynamics.
\\
-- The shock models largely overpredict the contribution of the 
photoionized precursor. Indeed, \cite{DS96} state that the column
density of ionized gas in the precursor might be overestimated
due to a possibly incorrect treatment of the transfer of the
UV ionizing photons (cf. end of Sect. 4.2 of \cite{DS96}).
%
% Modified
%
Moreover, the ionization structure of the
precursor could be much different than computed in \cite{DS96}  if the shock
evolves on time scales shorter than $\simeq$100 yr, i.e.
the recombination time in the pre--shock gas.

\begin{table}
\caption{Comparison between RCW103 (SNR), the Galactic center and the Circinus
galaxy}
\label{tab_cmpr}
\def\SKIP#1{\noalign{\vskip#1pt}}
\def\UNC{\rlap{:}}
\def\UNO{\rlap{$^{(1)}$}}
\def\DUE{\rlap{$^{(2)}$}}
\def\TRE{\rlap{$^{(3)}$}}
\def\NA{\rlap{$^a$}}
\def\NB{\rlap{$^b$}}
\def\NC{\rlap{$^c$}}
\def\ND{\rlap{$^d$}}
\def\NE{\rlap{$^e$}}
\def\X{ $\times$ }
\def\SB{$\rm [$}
\begin{flushleft}
\begin{tabular}{lccc}
\hline\hline
\SKIP{1}
 Line ratio & RCW103\UNO &   GC\DUE  & Circinus\TRE\\
\SKIP{2}
\hline
\SKIP{1}
%[FeII]\L26.0/[NeII]\L12.8 & 0.75 & 0.014 & 0.094 \\
%\SKIP{0}
[FeIII]\L22.9/[FeII]\L26.0 & 0.089 & 3.7 & -- \\
\SKIP{0}
[NeIII]\L15.6/[NeII]\L12.8 & 0.89 & 0.044 & 0.46 \\
\SKIP{0}
[NeV]\L14.3/[NeIII]\L15.6  & 0.012 & -- & 1.0 \\
\SKIP{0}
[NeIII]\L15.6/[FeII]\L26.0 & 3.1  & 1.2  & 5.4 \\
\SKIP{0}
[OIV]\L25.9/[FeII]\L26.0 & 0.30 & 0.24 & 8.8 \\
\SKIP{0}
[SiIX]\L3.94/[SiII]\L34.8  & $<$0.003 & -- & 0.025 \\
\SKIP{2}
\hline
\SKIP{2}
\end{tabular}
\def\NOTA#1#2{
\hbox{\vbox{\hbox{\hsize=0.030\hsize\vtop{\centerline{#1}}}}
      \vbox{\hbox{\hsize=0.97\hsize\vtop{\baselineskip4pt #2}}}}\vskip1pt}
\NOTA{ $^{(1)}$ }{ Line fluxes from this paper} 
\NOTA{ $^{(2)}$ }{ Data from Lutz et al. (\cite{galactic_center})}
\NOTA{ $^{(3)}$ }{ Data from Moorwood et al. (\cite{moorwood96})}
\end{flushleft}
\end{table}
\subsection{ Comparison with the Galactic center}

The region on the line of sight of the GC has a rich
spectrum of prominent IR lines which are believed to arise from gas
with an unusually large Fe gas phase abundance and
which is primarily photoionized by quite hot stars (Lutz et al.
\cite{galactic_center}).
Table~\ref{tab_cmpr} is a comparison between the most significant line ratios 
measured in RCW103 and in the GC.

The [FeIII]\L22.9/[FeII]\L26.0 ratio is much higher (a factor of 42) 
in the GC spectrum. This implies
Fe$^{++}$/Fe$^+\!>\! 1$ and a factor $\ge$10 larger than in RCW103,
regardless of the assumed gas density in the GC.
This simply reflects the fact that a region predominantly
photoionized by stars, such as those near to the GC, 
contains only a relatively small fraction of partially ionized gas. The 
recombining region behind the SNR shock front, on the contrary, has 
a large zone of partially ionized gas, which is heated by photoionization
from the shock front radiation, and where most of iron is forced into
Fe$^+$ by the very rapid charge exchange reactions with neutral hydrogen.

The [OIV]\L25.9/[FeII]\L26.0 ratio is the same in the two objects,
within the errors. Given the difficulties to produce
both FeII and OIV with photoionization from normal stars,
it seems not unreasonable to conclude
 that both species are primarily produced by shock excited gas
in the line of sight of the GC.

The [NeIII]\L15.6/[NeII]\L12.8 ratio is a factor of 20 lower in the GC than
in RCW103 
which indicates that fast shocks are more effective than
late O stars in producing NeIII.
Moreover, the [NeIII]\L15.6/[FeII]\L26.0 ratio is only a factor
of 2.6 higher in the GC than in RCW103 and this indicates that
a non negligible fraction of the [NeIII] emission from the GC could
come from shock excited gas.

\subsection{ Photoionized precursor and shocks in active galaxy nuclei }
\label{agn}

According to the shock models of \cite{DS96},
the precursor could be an important source of lines
from high ionization species (e.g. [OIII]\L5007), but its importance relative
to the post--shock region may strongly depend on the
column density of the pre--shock material. 
In a paper specifically dedicated to study
the spectral signatures of shocks in active galaxies,
Dopita \& Sutherland (\cite{dopita95}) consider the following
limiting cases: \\
-- Shock only, in which the precursor is very thin and its emission is
effectively negligible relative to the post--shock region. 
This can fairly well reproduce the line ratios observed in low excitation
AGNs (LINERS).\\
-- Shock~+~precursor, where the pre--shock region
is opaque to the ionizing photons from the shock front. 
Since the ionizing spectrum is quite hard and effectively similar
to a typical AGN spectrum, the ionization
structure of the precursor is similar to that of standard narrow
line regions photoionized by the AGN. Consequently, the emerging line spectrum
is similar to that of standard photoionization models and could explain,
therefore, the high excitation lines from e.g. type 2 Seyferts.

In view of this proposed scenario, it is interesting to compare the 
spectra of RCW103 with that of the Circinus galaxy, an archetype Seyfert 2
galaxy whose observed line ratios are listed in Table~\ref{tab_cmpr}.
The most striking difference is that the high excitation (coronal)
lines are much stronger in Circinus with, in particular, [NeV]/[NeIII]=1
and roughly 2 orders of magnitude larger than in RCW103.
Such a strong [NeV] could be in principle compatible with emission
from the precursor of a $v\!\sim\!500$ km/s shock (cf. Fig.~\ref{shock_pred}),
while even higher velocities, i.e. $\ga$1000 km/s, could probably account for 
highest ionization coronal lines (e.g. [SiX]).
The main problem is that these shocks should also emit prominent
low excitation lines from their fast moving post--shock gas, but this
is incompatible with the observed line profiles which are remarkably
narrow (FWHM$\le$150 km/s, Oliva et al. \cite{oliva94}) and similar
for all ionization species. Therefore, a shock dominated model for the Circinus
galaxy seems very unlikely and, more generally, the role played by the
photoionized precursor in Seyferts could be questioned on the basis
of the following arguments.

If dominated by photoionization,
the low excitation lines from the post--shock region (e.g. [SII])
should be broader than those from the photoionized precursor
(e.g. [OIII]), but this is in strict contrast with the observations 
which show that [OIII] and higher excitation lines are usually broader
than those of [SII] and lower excitation species. % (reference??).

The ISM medium of Seyfert galaxies is well known to be quite ``porous'',
especially within the ionization cones, and several arguments indicate
that the line emitting clouds are probably density bounded (e.g.
Binette et al. \cite{binette96}). The host galaxies of LINERS, on the
contrary, are often very rich in both gas and dust, a spectacular
example being NGC4945 (e.g. Moorwood et al. \cite{n4945}).
It seems therefore curious that the shocks in Seyferts should primarily
impact onto the relatively few large clouds (i.e. those with large
enough column density to absorb all the ionizing photons from the shock)
while, in LINERS, the shocks should selectively avoid the largest clouds
and only hit regions with low column densities (i.e. those which
cannot produce a bright precursor).

The absence of significant emission from the pre--shock region in 
RCW103 indicates
that shock models may overestimate the importance of the precursor
region. %, at least for shock velocities of about 300 km/s (cf. 
%Sect.~\ref{shock_models}).

\section{ Conclusions }

We have presented IR  spectroscopic observations of RCW103,
a relatively young and fast (about $10^3$ yr and $\simeq$1200 km/s, 
Nugent et al. \cite{nugent}) supernova remnant whose prominent line
emitting filaments result from the interaction of the fast  SNR blast wave
with ISM clouds.
The secondary shocks driven into the higher density medium have
velocities $\ga$300 km/s, as indicated by the observed dynamics of the
filaments (cf. Fig.\ref{velcuts}), by the very high
surface brightness of the lines and by velocity sensitive
line ratios such as [NeV]/[NeIII] (Sect.~\ref{shock_models}).

The spectrum is dominated by prominent lines from low excitation species
which have been used to estimate the density, temperature and abundances
of the emitting region. The results indicate this is post--shock gas with 
relatively low density ($\Ne\!\sim\!10^3$ cm$^{-3}$), normal ISM abundance
and essentially free of dust. Although the [NiII] 
lines seem to require a large  
nickel overabundance we argue, however, that this simply reflects 
uncertainties in the atomic parameters and propose that the collision
strength for [NiII]\L6.64 is $\Upsilon$=1 and a factor of 7
larger than the computed value. 
Smaller uncertainties in some of the atomic
parameters for [FeII] are also suggested by the data which nevertheless
exclude large errors in the computed collision strengths.

An important discrepancy with fast shock models is that
the prominent high excitation lines predicted from the photoionized
precursor are not seen in RCW103.
Since low column densities in the pre--shock region are ruled out
by the observed H$_2$ emission from the shock precursor, this
discrepancy probably indicates that shock models largely overestimate
the strength of the lines from the precursor. This may cause
problems to shock models of Seyfert galaxies where the high excitation
lines should be dominated by emission from the photoionized pre--shock region.

\begin{acknowledgements}
E. Oliva acknowledges the partial support of the Italian Space Agency (ASI)
through the grant ARS--98--116/22.
We thank Henrik Spoon for help with the ISOPHOT-S data. SWS and the ISO
Spectrometer Data Center at MPE are supported by DLR (formerly DARA) under
grants 50--QI--8610--8 and 50--QI--9402--3.
\end{acknowledgements}


\begin{thebibliography}{}

\bibitem[1989]{arendt}        %OK
Arendt R.G., 1989, ApJS 70, 1

%\bibitem[1996]{bautista_pradhan}
%Bautista M.A., Pradhan A.K., 1996, A\&AS 115, 551

\bibitem[1996]{bautista}        %OK
Bautista M.A., Peng J., Pradhan A.K., 1996, ApJ 460, 372

\bibitem[1996]{binette96}        %OK
Binette L., Wilson A. S., Storchi-Bergmann T., 1996, A\&A 312, 365

\bibitem[1983]{leibowitz}        %OK
Leibowitz E.M., Danziger I.J., 1983, MNRAS 204, 273

\bibitem[1986]{dennefeld}        %OK
Dennefeld M., 1986, A\&A 157, 267

\bibitem[1995]{dopita95}         %OK
Dopita M.A., Sutherland R.S., 1995, ApJ 455, 468 

\bibitem[DS96]{DS96}         %OK
Dopita M.A., Sutherland R.S., 1996, ApJS 102, 161 (DS96)

\bibitem[1993]{drainemckee}         %OK
Draine B.T., McKee C.F., 1993, ARA\&A 31, 373 

\bibitem[1987]{hester}   %OK
Hester J.J., 1987, ApJ 314, 187

\bibitem[1989]{hollenbach89}   %OK
Hollenbach D., McKee C.F., 1989, ApJ 342, 306

\bibitem[1996]{liu}         %OK
Liu X.W. et al., 1996, A\&A 315, L257

\bibitem[1995]{lucy}         %OK
Lucy L.B., 1995, A\&A 294, 555

\bibitem[1996]{galactic_center}         %OK
Lutz D. et al., 1996, A\&A 315, L269

\bibitem[1987]{moorwood87}         %OK
Moorwood A.F.M., Danziger I.J., Oliva E., 1987, The Messenger 48, 49

\bibitem[1996]{n4945}         %OK
Moorwood A.F.M., van der Werf P.P., Kotilainen J.K., Marconi A.,
Oliva E., 1996
A\&A {\bf 308}, L1 

\bibitem[1996]{moorwood96}         %OK
Moorwood A.F.M., Lutz D., Oliva E., Marconi A., Netzer H.,
 Genzel R., Sturm E., de Graauw T., 1996, A\&A 315, L109 

\bibitem[1984]{nugent}         %OK
Nugent J.J., Pravdo S.H., Garmire G.P., Becker R.H., Tuohy I.R., Winkler P.F,
1984, ApJ 284, 612

\bibitem[1988]{nussbaumer88}         %OK
Nussbaumer H., Storey P.J., 1988, A\&A 193, 333

\bibitem[OMD89]{OMD89}         %OK
Oliva E., Moorwood A.F.M., Danziger I.J., 1989, A\&A 214, 307 (OMD89)

\bibitem[OMD90]{OMD90}         %OK
Oliva E., Moorwood A.F.M., Danziger I.J., 1989, A\&A 240, 453 (OMD90)

\bibitem[1994]{oliva94}         %OK
Oliva E., Salvati M., Moorwood A.F.M., Marconi A., 1994, A\&A 288, 457 

%\bibitem[1989]{osterbrock89}
%Osterbrock D.E., 1989, Astrophysics of Gaseous Nebulae and Active Galactic 
%Nuclei, University Science Books, New York

\bibitem[1993]{pradhan_zhang}         %OK
Pradhan A.K., Zhang H.L., 1993, ApJL 409, L77

\bibitem[1996]{quinet}         %OK
Quinet P., Le Dourneuf M., Zeippen C.J., 1996, A\&AS 120, 361        %OK

\bibitem[1979]{raymond}        %OK
Raymond J.C., 1979, ApJS 39, 1

\bibitem[1990]{russel}        %OK
Russel S.C., Dopita M.A., 1990, ApJS 74, 93

\bibitem[1979]{shull_mckee}        %OK
Shull J.M., McKee C.F., 1979, ApJ 227, 131

\end{thebibliography}
\end{document}